\documentclass[9pt,twocolumn,twoside,superscriptaddress]{revtex4-2}

\usepackage[english]{babel}
\usepackage[utf8]{inputenc}
\usepackage[T1]{fontenc}

\usepackage{graphicx}
\usepackage{verbatim}
\usepackage{amsmath}
\usepackage{amssymb}
\usepackage{mathrsfs}
\usepackage{hyperref}
\usepackage[title]{appendix}
\usepackage{mathtools} 
\usepackage{graphicx}
\usepackage{xcolor}
\usepackage{upgreek}

\usepackage{array}

\usepackage{chemmacros}

\graphicspath{{figures/}}

\newcommand{\be}{\begin{equation}}
\newcommand{\ee}{\end{equation}}
\newcommand{\DGh}{\Delta G_{\rm hyb}}

\begin{document}
\title{Emergence of Homochirality via Template-Directed Ligation in an RNA Reactor}

\author{Gabin Laurent}
\affiliation{Physics of Complex Biosystems, Technical University of Munich, 85748 Garching, Germany}
\affiliation{Gulliver UMR CNRS 7083, ESPCI Paris, Universit\'e PSL, 75005 Paris, France}
\affiliation{Univ. Grenoble Alpes, CNRS, UMR 5525, VetAgro Sup, Grenoble INP, TIMC, 38000 Grenoble, France}
\author{Tobias Göppel}
\affiliation{Physics of Complex Biosystems, Technical University of Munich, 85748 Garching, Germany}
\author{David Lacoste}
\affiliation{Gulliver UMR CNRS 7083, ESPCI Paris, Universit\'e PSL, 75005 Paris, France}
\author{Ulrich Gerland}
\email[]{gerland@tum.de}
\affiliation{Physics of Complex Biosystems, Technical University of Munich, 85748 Garching, Germany}

\begin{abstract}
RNA in extant biological systems is homochiral; it consists exclusively of \iupac{\D-ribonucleotides} rather than \iupac{\L-ribonucleotides}. How the homochirality of RNA emerged is not known. Here, we use stochastic simulations to quantitatively explore the conditions for RNA homochirality to emerge in the prebiotic scenario of an `RNA reactor', in which RNA strands react in a non-equilibrium environment. These reactions include the hybridization, dehybridization, template-directed ligation, and cleavage of RNA strands. The RNA reactor is either closed, with a finite pool of ribonucleotide monomers of both chiralities (\iupac{\D} and \iupac{\L}), or the reactor is open, with a constant inflow of a racemic mixture of monomers. For the closed reactor, we also consider the interconversion between \iupac{\D} and \iupac{\L} monomers via a racemization reaction. 
We first show that template-free polymerization is unable to reach a high degree of homochirality, due to the lack of autocatalytic amplification. In contrast, in the presence of template-directed ligation, with base pairing and stacking between bases of the same chirality thermodynamically favored, a high degree of homochirality can arise and be maintained provided the non-equilibrium environment overcomes product inhibition, for instance via temperature cycling. Indeed, if the experimentally observed kinetic stalling of ligation after chiral mismatches is also incorporated, the RNA reactor can evolve towards a fully homochiral state, in which one chirality is entirely lost. This is possible because the kinetic stalling after chiral mismatches effectively implements a chiral cross-inhibition process. 
Taken together, our model supports a scenario, where the emergence of homochirality is assisted by template-directed ligation and polymerization in a non-equilibrium RNA reactor. 
\end{abstract}

\maketitle

\section{Introduction}
Life on earth relies mainly on chiral molecules, which are, by definition, molecules not superposable to their mirror images. The  observation that chiral molecules forming biopolymers (such as proteins, DNA or RNA) are present under only one of their two possible forms (\iupac{\D} or \iupac{\L} enantiomers) in living matter is referred to as homochirality or biological asymmetry \cite{meierhenrich_amino_2008}.
Spontaneous mirror symmetry breaking (SMSB) is a possible hypothesis used to understand biological homochirality. In an alternative hypothesis, a small initial bias causes an explicit symmetry breaking (ESB) \cite{sallembien_possible_2022, blanco_chiral_2011, saito_chirality_2005}. One possible mechanism for ESB is the interaction of prebiotic molecules with circularly polarized light, which could explain the bias observed in amino acids on comets \cite{meinert_amino_2022}, while another proposal is that the parity violation of the weak force creates a small difference in thermodynamic properties between two enantiomers \cite{globus_chiral_2020}. 
 
Whether biological homochirality proceeds via SMSB or ESB and whatever the cause of the initial chiral bias may be, a mechanism of  amplification of chiral biases is needed to explain the full homochirality observed in biological systems. The classic paper by Frank presented a simple mathematical model based on autocatalytic reactions, which provides such an amplification mechanism \cite{frank_spontaneous_1953}. An important ingredient in this model, besides autocatalysis, is the presence of a reaction of mutual antagonism between the two enantiomers (chiral inhibition).
Despite its importance, the Frank model is too simple to describe prebiotically relevant  systems because this model considers only a few molecular species (an achiral one and two chiral ones) and very specific reactions, 
whereas prebiotic chemistry likely involved a large number of species and reactions.
Recently, two of us introduced a model for the emergence of homochirality using ideas from random matrix theory, which precisely takes into account that prebiotic chemistry is messy, with many reactions and species involved  \cite{laurent_emergence_2021}. This approach shows that homochirality can emerge in autocatalytic chemical networks in a robust way, as long as the network is large enough (in terms of the number of its chiral species and reactions) and driven sufficiently far from equilibrium.

Polymerization produces a large number of different molecular species as the length of the polymers increases, and the question of whether and how polymerization can support the emergence of homochirality arises naturally. Links between polymerization and homochirality have been considered in theoretical models \cite{sandars_toy_2003, brandenburg_homochiral_2005, brandenburg_dissociation_2005, gleiser_extended_2008, wu_autocatalytic_2012}, underlining the importance of enantioselective autocatalysis and chiral inhibition. On the experimental side, Viedma deracemization experiments  have demonstrated that full homochirality in a solid phase can be obtained using a proteinogenic amino acid as monomers \cite{viedma_evolution_2008}. The highest selective enrichment of one enantiomer is observed with the aid of particle grinding. Particle grinding puts a large number of monomers in solution, which strongly favors polymerization over depolymerization. Then, Ostwald ripening, leads to a competition between the two chiralities \cite{blanco_viedma_2013}. This technique has been used to achieve the selective production of pharmaceutical compounds with a desired chirality \cite{noorduin_emergence_2008}.

\begin{figure*}
	\includegraphics[width=0.72\linewidth]{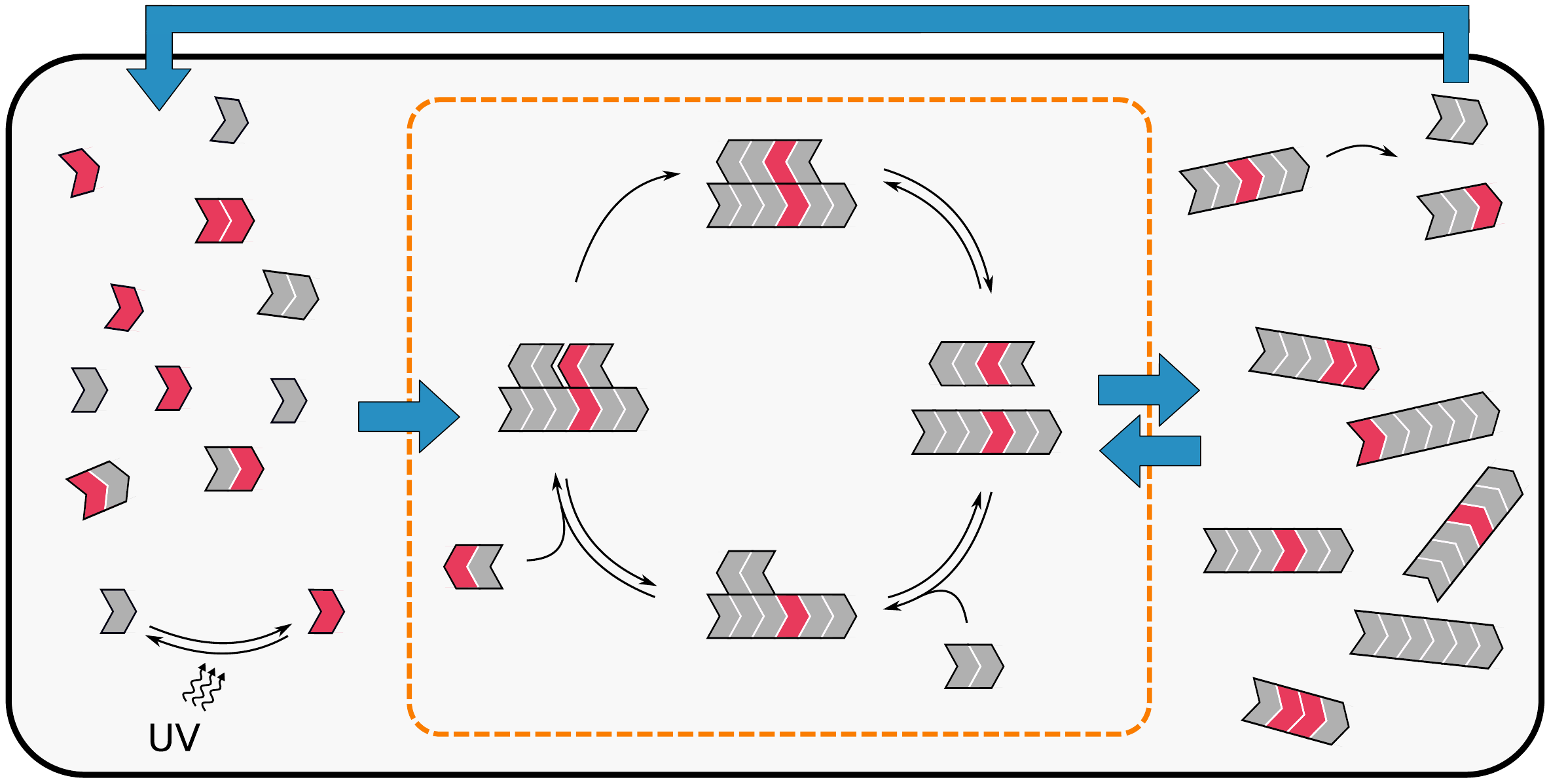}
	\caption{Global polymerization scheme directed by templates. Gray and red colors represent the \iupac{\D} and \iupac{\L} chiralities, respectively. Starting from an initial racemic pool of monomers and dimers, the system gets enriched in longer polymers with a preferred chirality, due to an autocatalytic cycle based on templated ligation, which is represented by the central part of the figure inside the orange dashed line. Outside this area on the left, the racemization of monomers due to exposure to UV light is represented. The two arrows on the right-hand side of the autocatalytic cycle represent long strands that have been produced by the autocatalytic cycle now serving as templates in the cycle, with the upper arrow representing the recycling of monomers and dimers, produced by the hydrolysis of single-stranded polymers present in the system.}
	\label{fig:global_scheme}
\end{figure*}

While these works help in the understanding of general conditions for achieving homochirality, they do not apply to template-directed polymerization of RNA, which needs to be considered to link homochirality to RNA world scenarios for the emergence of life. 
A significant advantage of template-directed polymerization over simple chain-growth or step-growth polymerization is that it immediately satisfies the requirement of autocatalysis. Indeed, the ligation between two strands on an RNA template represents an autocatalytic step, which can be thermodynamically selected via pairing interactions between the template and the ligated strands. Experiments show that template-directed polymerization is inherently chiroselective and can be achieved without enzymes \cite{joyce_chiral_1984, bolli_pyranosyl-rna_1997}. Thus this system appears to display all features required to produce a transition to homochirality under prebiotic conditions. Previous computational analyses indeed favor this conclusion. 
Tupper {\it et al.} \cite{tupper_role_2017} introduced a kinetic model that was designed to study the kinetic competition between a simple polymerization process without template assistance and an idealized template-directed ligation process. For that purpose, the authors assumed that template-directed ligation does not allow for any chiral mismatches, such that template-directed ligation by itself is guaranteed to conserve homochirality. The authors then identified a transition, in which the chiral symmetry is spontaneously broken as a function of the ratio of the kinetic rate constants associated with the two kinetic processes.        
Chen and Ma \cite{chen_origin_2020} simulated a different model, which includes nucleotide precursors of both chirality and  combines template-directed synthesis with surface-mediated synthesis of single strands. In their model, a strong form of chiral inhibition is assumed for both polymerization processes: A monomer of the wrong chirality acts as a chain terminator in surface-mediated synthesis and only monomers of the same chirality as the RNA template can be incorporated into extending chains. The authors reported computational experiments, in which the assumed mechanisms generate a strong chiral bias, but not full homochirality. 

Here we focus primarily on the template-directed ligation process, and ask under what conditions this process can lead to full homochirality within a non-equilibrium RNA reactor, starting from a racemic mixture of short oligomers. Importantly, our model of template-directed ligation is based on a thermodynamically consistent kinetic description of RNA hybridization, with finite free energy penalties for chiral mismatches in a hybridized RNA complex (for mismatches both along a strand and across strands). 
Our model, which is schematically illustrated in Fig.~\ref{fig:global_scheme}, derives from the RNA reactor model of Ref.~\cite{goppel_thermodynamic_2022}. As such, it describes the kinetic interplay of RNA hybridization, ligation, and cleavage (due to hydrolysis) within an environment that is both chemically and physically out of equilibrium, due to chemical activation of nucleotides and periodically varying ambient temperature. As opposed to Ref.~\cite{goppel_thermodynamic_2022}, the present model introduces nucleotides with different chiralities. Since it allows for chiral mismatches, it explicitly describes the kinetic competition between perfectly homochiral strands and the entropically favored, but energetically disfavored, strands consisting of nucleotides with mixed chiralities. As we do not focus on sequence-dependent effects here, our model uses only nucleotides with a single type of base that is assumed to be self-complementary. This simplifies the simulations and the analysis.

This article is organized as follows. In Sec.~\ref{sec: non-template polymerization}, we first study non-templated polymerization, which we find unable to achieve a high degree of homochirality even in the most favorable conditions. In Sec. ~\ref{model_section}, we then turn to the study of template-directed polymerization. We present simulation data in Sec.~\ref{sec:Results} and discuss which conditions yield optimal results. We interpret our simulation results in Sec.~\ref{sec:discussion}. In Sec.~\ref{sec:conclusion}, we provide a summary and discuss our conclusions.

\section{Template-free polymerization}
\label{sec: non-template polymerization}
\subsection{A simple model for step-growth polymerization}
\begin{figure}
\includegraphics[width=0.99\linewidth]{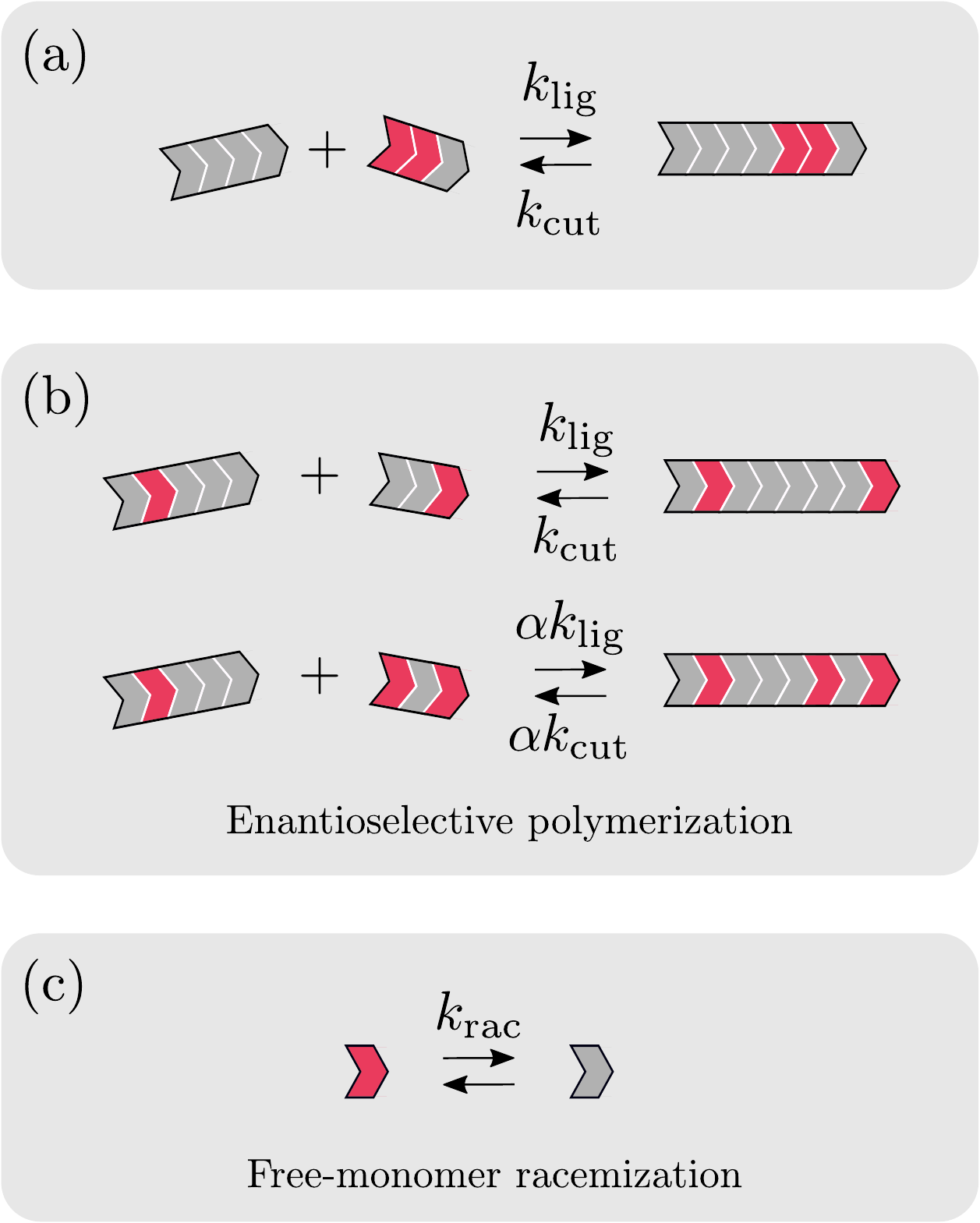}
\caption{Step-growth polymerization. Gray and red colors represent the \iupac{\D} and \iupac{\L} chiralities. (a) Template-free polymerization in a chiral system. (b) Step-growth polymerization is slowed down by a factor $\alpha$ as a result of the incorporation of a monomer of the wrong chirality due to the interaction with a mineral surface such as montmorillonite clay. (c) Racemization of free nucleotides.}
\label{fig:step_growth_panel}
\end{figure}

Let us first consider step-growth polymerization [as represented in Fig.~\ref{fig:step_growth_panel}(a)]
as a reference model.
The system is initially composed of two enantiomeric families of monomers and oligomers 
 that can assemble together with a rate constant $k_{\rm lig}$. Any single strand can be cleaved by hydrolysis with a rate constant $k_{\rm cut}$ for every phosphodiester bond in the  strand, i.e. a linear strand of length $L$ has the rate $(L-1)k_{\rm cut}$ to be cleaved at one of its bonds. 
Without catalysis, there is no evidence for a chiroselective behavior during step-growth polymerization of RNA, and it is generally assumed that in a racemic pool of monomers, strands composed of nucleotides of random chiralities will be produced \cite{joshi_progress_2011}. However, this process can become chiroselective when performed on a mineral surface such as montmorillonite clay \cite{joshi_selectivity_2007, joshi_progress_2011}. 
In this case, the ligation of two oligomers or monomers is slowed by a factor $\alpha < 1$ when the ligation occurs between two nucleotides of distinct chiralities (and only in that case), i.e., the ligation rate is reduced to the value $k'_{\rm lig} = \alpha k_{\rm lig}$ as depicted in Fig.~\ref{fig:step_growth_panel}(b). In our model, the cleavage rate for bonds between two nucleotides of different chiralities is also reduced to $k'_{\rm cut} = \alpha k_{\rm cut}$ such that the same equilibrium constant is obtained as for bonds between nucleotides of the same chirality: $K=k_{\rm lig}/k_{\rm cut}= k'_{\rm lig}/ k'_{\rm cut}$.

We performed stochastic simulations of this system using the Gillespie algorithm, in a closed reactor. When it is explicitly mentioned, we include a racemization reaction at the monomer level in this system as shown in Fig.~\ref{fig:step_growth_panel}(c). This reaction is a one-step process that converts \iupac{\D} monomers into \iupac{\L} monomers and vice versa. The reason for this additional reaction is that it allows to break the conservation law of \iupac{\D} monomers and \iupac{\L} monomers, which would otherwise prevent the system from reaching a fully homochiral state. At the same time, it does so while keeping the system closed and well mixed. In Sec.~\ref{subsec:open_reactor}, we relax this condition and explore the case of open systems without racemization reactions. Experimentally, racemization reactions are known to occur when chiral molecules are exposed to UV light in a process known as photoracemization \cite{belletti_photoracemizationbased_2020}. 
While the conversion of nucleotides cannot occur in a single-step reaction, we nevertheless treat this conversion as a single step for simplicity.

\begin{figure*}
	\centering
	\includegraphics[scale=0.6]{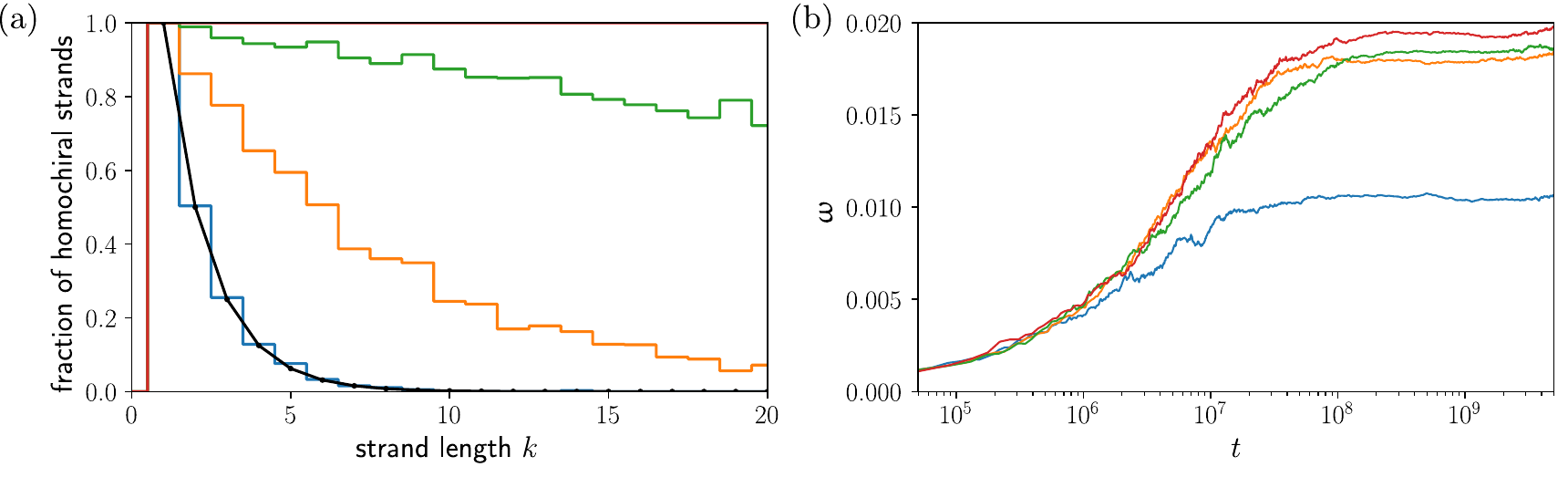}
	\caption{Stochastic simulations of step-growth polymerization. (a) Fraction of homochiral strands as a function of their length $k$ in the steady state, without any racemization reactions. The black curve is the theoretical prediction for unbiased homochiral and heterochiral ligation. In this case, the fraction of homochiral strands is $2^{1-k}$. (b) Time evolution of the enantiomeric excess at the monomer level. The simulations were performed with instantaneous racemization reactions for monomers that are not polymerized. Here $\alpha$ is the ligation chiral bias $\alpha = k'_{\rm lig}/k_{\rm lig}$. Data are averaged over 100 independent numerical realizations starting from an initial racemic pool composed of 2500 \iupac{\D} and 2500 \iupac{\L} monomers. The parameters are $k_{\rm lig} = 5.27\times 10^{-9}$, $k_{\rm cut} = 10^{-10}$, $V = 4.69 \times 10^{-4} \upmu $m$^3$ and $\alpha = 1$ (blue curve), $\alpha = 0.1$ (orange curve), $\alpha = 0.01$ (green curve), and $\alpha = 10^{-10}$ (red curve).}
	\label{random_polymerization}
\end{figure*}

\subsection{Simulation results}
In the absence of racemization reactions, we find that the distribution of homochiral strands as a function of the length of the strand produced in the reactor shows a deviation from the random case in the presence of a chiroselective catalyst [see Fig.~\ref{random_polymerization}(a)]. The system reaches a state where all strands taken individually are homochiral in the case of perfect chiral selection ($\alpha = 0$), while the pool remains generally racemic.

Let us define the enantiomeric excess $\upomega$ as the absolute value of the difference between the total number of monomers of one chirality minus that of the other chirality, normalized by their sum, taking into account all monomers in the system, the polymerized and non polymerized ones. 
In the presence of fast racemization reactions at the monomer level (the pool at the free monomer level is racemic at any time), we observe in Fig.~\ref{random_polymerization}(b) that a bias in the enantiomeric excess of the order of $\upomega \sim 1\%$ is present in the steady state even in the absence of chiral selection ($\alpha = 1$). 
This bias is due to a finite-size effect, i.e., it is caused by the finite total number of monomers present in the simulation: The bias tends to zero as the size of the nucleotide pool is increased (see Fig. S1 in the Supplemental
Material).

When a selection is present, the system ends up in a steady state with a slightly larger $\upomega$ that hardly reaches $2\%$ in the most favorable case of perfect chiral selection.
In addition, this bias is highly dependent on the presence of the chiroselective catalyst. If the catalyst were to disappear from the system, the $\upomega$ would relax to the small steady-state value observed when no chiral selection is present. 

To summarize, although catalyzed step-growth polymerization can favor homochiral strands due to chiral enzymes or the contact with mineral surfaces, 
no significant chiral bias can be produced by this system. We attribute this failure of simple polymerization to the lack of an autocatalytic amplification mechanism. 
To overcome these limitations, we now consider template-directed polymerization.

\section{Model of Template-directed polymerization}
\label{model_section}
\begin{figure}[h!]
\includegraphics[width=0.99\linewidth]{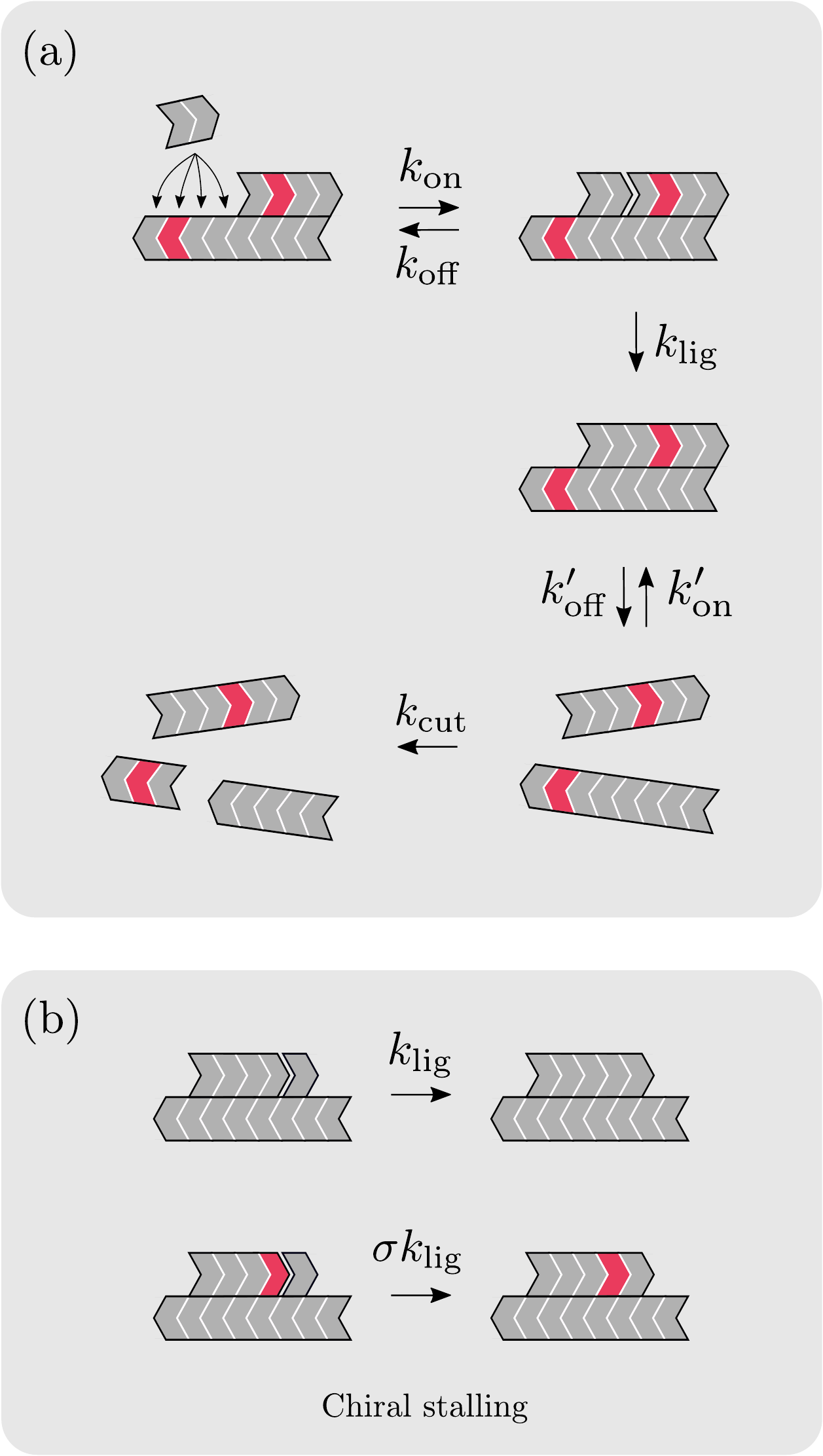}
\caption{Templated-ligation schemes. Gray and red colors represent \iupac{\D} and \iupac{\L} chiralities, respectively. (a) Template-directed ligation of activated nucleotides. The four conjoined arrows depict the four possibilities for the dimer to bind to the longer template. The subsequent step of ligation reaction with rate $k_{\rm lig}$ and separation of the strands in the complex are also represented with their rates. (b) The ligation of two strands on a template is slowed by a factor $\sigma$ if at the ligation site, there is a chirality mismatch between one of the two ligated strands and the template.}
\label{fig:templated_ligation_panel}
\end{figure}

We now consider a chemical system where, starting from a pool of nucleotides, template-directed ligation involving RNA polymers and monomers occurs, as represented in Fig.~\ref{fig:templated_ligation_panel}(a). We assume the required activation chemistry for monomers is present in the system.

\subsection{Hybridization and dehybridization}
Hybridization and dehybridization of two strands are assumed to be single-step processes. Two complexes collide with a rate $k_{\rm coll}$ and hybridize with equal probabilities for all the possible binding configurations. In practice, $k_{\rm coll}$ is neither calculated nor estimated; instead its value is used as reference to compute the other rates as explained in Table~\ref{tab:parameter_table}. The rates of hybridization and dehybridization, $k_{\rm on}$ and $k_{\rm off}$, are constrained by the thermodynamic relation \cite{dill_molecular_2010, rosenberger_self-assembly_2021}
\begin{equation}
\frac{k_{\rm off}}{k_{\rm on}} = (V N_{\rm A} c^{\circ}) \exp \left(\frac{\DGh}{k_B T} \right) \, ,
\label{thermo_consist}
\end{equation}
where $V$ is the reactor volume, $N_{\rm A}$ the Avogadro number, $c^{\circ} = 1$ mol/l the reference concentration, $k_B$ the Boltzmann constant, $T$ the temperature, and $\DGh$ the hybridization energy associated with the hybridized part of a duplex.
The hybridization rate $k_{\rm on}$ between two strands for one specific combination is related to the constant collision rate $k_{\rm coll}$ through the relation
\be
k_{\rm on} = \frac{1}{\theta} k_{\rm coll} \, ,
\ee
where $\theta$ is the number of possible duplex combinations they can form. Two strands of length $l_1$ and $l_2$ have $\theta = l_1 + l_2 - 1$ possible ways to attach to each other through base paring. This expression changes if one of the strands is already part of a complex, but the total rate for the hybridization of two strands still sum to $k_{\rm coll}$. The computation of $\Delta G_{\rm hyb}$ depends on the choice of an energy model.
In the first place, it is important to note that $\Delta G_{\rm hyb}$ does not depend on just the length of the hybridized part within a complex. As base-pair mismatches in DNA and RNA destabilize a complex, chiral mismatches also destabilize \cite{damha_oligodeoxynucleotides_1991, damha_antisense_1994, urata_thermodynamic_2003, kawakami_thermodynamic_2005, hauser_utilising_2006, szabat_thermodynamic_2016} heterochiral DNA and RNA complexes.

A simple way to evaluate the hybridization energy of a complex would be to sum all contributions for all base pairs in the duplex (a duplex being a complex composed of two hybridized strands). In practice, this method is not accurate enough,
because it is crucial to consider neighbor interactions within each complex strand \citep{liu_homochirality_2020}. Indeed, with a base-pair model, a purely alternating duplex with homochiral base pairs would have the same Gibbs free energy and therefore the same stability as the associated fully homochiral duplex, in contradiction with experimental observations \cite{hauser_utilising_2006}. Therefore, a block wise model is needed to describe correctly the thermodynamics of hybridization  \citep{turner_nndb_2010}.
\begin{figure}[h]
	\includegraphics[scale=0.35]{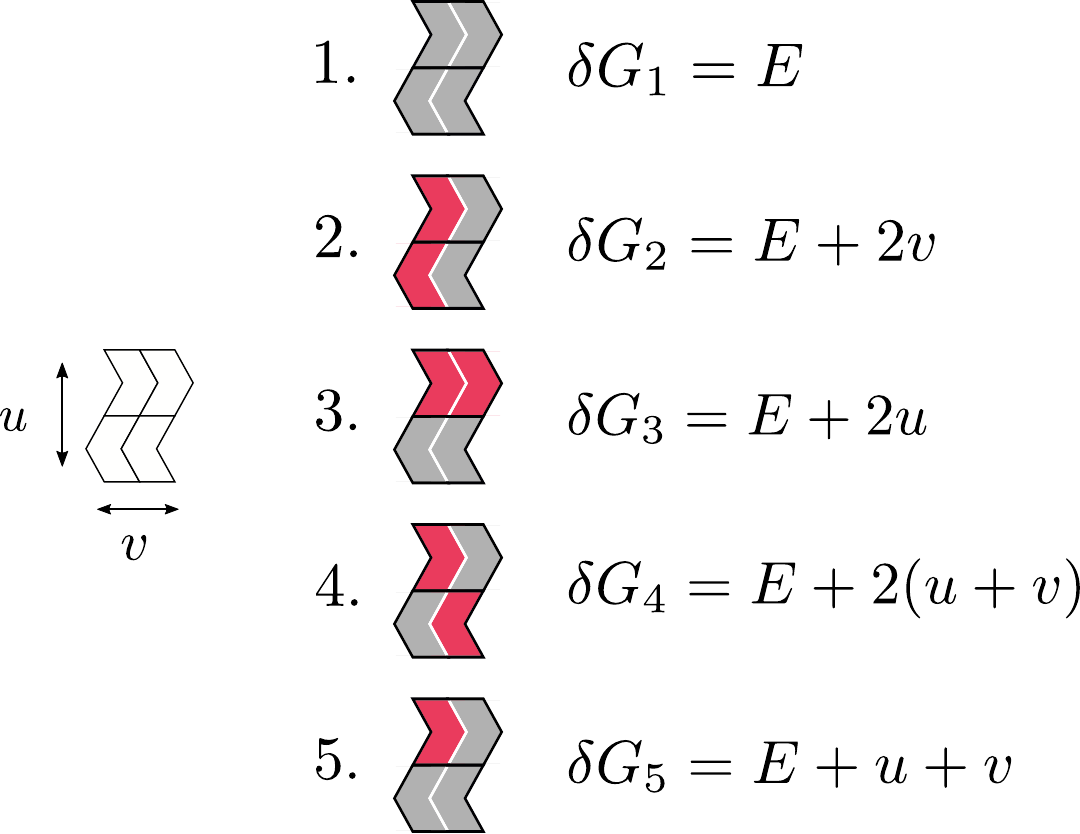}
	\caption{The five different kinds of four-monomer blocks and their associated free energies as a function of the basal free energy of a fully homochiral block, $E$, and the vertical and lateral penalties $u$ and $v$ due to chiral mispairing or heterochiral patterns within the two distinct strands.}
	\label{block_wise_model}
\end{figure}
The block wise energy model takes into account the contributions of blocks of two consecutive base pairs rather than individual base pairs. It accounts for base pair and lateral interactions within each strand, when a duplex structure is formed during hybridization. The total hybridization energy is summed over all blocks in the complex
\begin{equation}
\DGh = \sum_{i \in \{ \rm blocks\}} \delta G_i \, ,
\end{equation}
where $\delta G_i$ is the free-energy contribution of the $i$th block. Turner and Mathews \cite{turner_nndb_2010} and SantaLucia and Hicks \cite{santalucia_thermodynamics_2004} used experimental data to infer all hybridization block energy contributions for every possible base-pair combination for \iupac{\D-RNA} or \iupac{\D-DNA} in a duplex, at $T = 37 \, ^{\circ}{\rm C}$. Unfortunately, block energies have not been determined extensively for DNA and RNA complexes of mixed chiralities. One can however circumvent the lack of experimental data by building a minimal model to describe heterochiral interactions within a block of four monomers (see Fig.~\ref{block_wise_model}).
We define as $\delta G_1 = E < 0$ the hybridization block energy associated with a fully homochiral block. Averaging over all blocks from the NNDB \cite{turner_nndb_2010} (for RNA), we set $E=-1.59$~kcal.mol$^{-1}$. We denote by $u$ the vertical penalty associated with a mismatch of chiralities in a base pair, and $v$ is the lateral penalty associated with heterochiral patterns in one of the strands of the complex ($u$ and $v$ are positive quantities). The block energies are thus ordered in amplitude either as $\delta G_1 <  \delta G_2 \leq \delta G_5 \leq  \delta G_3 < \delta G_4$ or as $\delta G_1 <  \delta G_3 \leq \delta G_5 \leq  \delta G_2 < \delta G_4$.
The double diagonal mismatch destabilizes most of the complex due to an incorrect base pairing and incompatible rotational twists in the duplex that is supposed to form a helix. We differentiate three different cases depending on the relative values of $u$ and $v$ ($u < v$, $ v< u$, and $ u \sim v$) describing the strength of lateral or vertical mismatch penalty. From the experimental literature \cite{urata_thermodynamic_2003}, we deduce an approximate value for $u+v$ to be $u+v \sim 1.26$~kcal.mol$^{-1}$. The three different couples of values for $(u,v)$ are $(0.42,0.84)$, $(0.84,0.42)$, and $(0.63,0.63)$~kcal.mol$^{-1}$. For the case of terminal blocks formed out of three monomers instead of four, when the end of a strand does not coincide with the end of the template in a complex, we use the same model but with a different energy for the homochiral three-monomer block: $E_{\rm 3-block}=-0.47$~kcal.mol$^{-1}$ (this value is also the result of an average of the data from the NNDB for RNA). In the simulations, all block energies and penalties are rescaled by a factor 0.5 to fit with the parameters of Table~\ref{tab:parameter_table}. 

\subsection{Ligation and stalling}
In this model, we neglect template-free ligation, i.e. ligation occurring without templates, as it is slower than the templated ligation. Two strands can ligate on a template with the rate constant $k_{\rm lig}$ if they are bound adjacently on the template [see Fig.~\ref{fig:templated_ligation_panel}(a)]. During the template-directed ligation of RNA, the ligation speed depends on the chiral structure near the ligation site. Experimental evidence \citep{joyce_chiral_1984, bolli_pyranosyl-rna_1997} suggests that the ligation of two strands on a template is slowed when the chirality of the nucleotide at the ligated end of a strand does not match the chirality of the template [see Fig.~\ref{fig:templated_ligation_panel}(b)]. This stalling slows the ligation by one to two orders of magnitude \cite{joyce_chiral_1984}. For homochiral binding sites, the ligation occurs at its basal rate $k_{\rm lig} = \lambda$. Therefore, we assume that the ligation rate for a given complex is described by
\begin{equation}
k_{\rm lig} = \left\{
\begin{array}{ll}
	\lambda \, \sigma^2 & \mbox{if } C_S^{-1} \neq C_T^{-1}, \,\, C_S^{+1} \neq C_T^{+1} \\
	\lambda \,  \sigma & \mbox{if } C_S^{\pm 1} \neq C_T^{\pm 1}, \,\, C_S^{\mp 1} = C_T^{\mp 1}\\
	\lambda & \mbox{otherwise.}
\end{array}
\right.
\end{equation}
where $C_{S}^i$ ($C_{T}^i$) describes the chirality of the {\it i}th monomer of a strand participating in a ligation event (the template strand), which can be either \iupac{\D} or \iupac{\L}, and $\sigma \leq 1$ is the stalling factor. Monomers on both sides of the ligation site are labeled $-1$ and $+1$. 
This kinetic phenomenon (chiral stalling) induces a powerful chiroselective effect during template-assisted RNA ligation. Note that chiral stalling is fundamentally different from an energetically favored formation of chiral homodimers over heterodimers.

\subsection{Hydrolysis and temperature cycles}
Hydrolysis can cleave RNA strands, by breaking a phosphodiester bond in the RNA backbone. We denote the cleavage rate by $k_{\rm cut}$ and assume it is the same for all phosphodiester bonds in the system. Therefore, any strand of length $L$ is susceptible to be cleaved at one of its junctions with a rate $(L-1)k_{\rm cut}$, resulting in two shorter strands. We assume that only single-strand RNA can be cleaved because it has been shown that the duplex formation of RNA impedes the hydrolysis of phosphodiester bonds \cite{usher_rna_1972, zhang_duplex_2021}.

From the thermodynamic relation (\ref{thermo_consist}), it appears that the hybridization energy of long complexes becomes arbitrarily large in negative values and results in frozen complexes that do not dehybridize in a finite time. To circumvent this well-known issue, known as product inhibition, which would lead to frozen dynamics with all strands present in highly stable complexes, we impose a thermal cycling to the system which melts all complexes periodically irrespective of their length \cite{mast_thermal_2010, salditt_thermal_2020}. In this thermal cycling, instantaneous temperature peaks occur  periodically with a given frequency $1/ \tau_{\rm cycle}$ and split all complexes into their single-stranded components. We assume here that these temperature peaks occur quickly enough to not affect the other processes in the system even though, in principle, ligation and hydrolysis are also affected by the temperature change.

\section{Template-directed polymerization: results}
\label{sec:Results}
\subsection{Spontaneous chiral symmetry breaking in a closed reactor}
\label{sec:results_closed_reactor}
We now study the model of template-directed ligation introduced above by simulating its resulting behavior in a closed reactor using a Gillespie algorithm, as previously described \cite{goppel_thermodynamic_2022}. Briefly, the variety of molecular species and complexes that can appear in the system is not fixed in advance. Instead, the molecular inventory of the reactor is dynamically updated, adding (removing) species and complexes that newly appear (disappear) after a given reaction. In addition, the list of all possible reactions in the system and the associated rate constants is updated after each reaction event.
The system is initially inoculated with a total of 5000 nucleotides distributed in 4920 monomers (2460 \iupac{\D} and 2460 \iupac{\L} monomers) and 40 dimers (10 \iupac{\D\D}, 10 \iupac{\L\L}, 10 \iupac{\D\L} and 10 \iupac{\L\D} dimers) in a racemic fashion. Chiral biases will emerge naturally as a consequence of the stochasticity of the simulation. All parameters used for the simulations are given in Table~\ref{tab:parameter_table}. 

Simulation data indicate that there is no significant difference in the dynamics between the three different parameter sets for the block energy model ($u < v, v < u$ or $u \sim v$). For every quantity studied here, the dynamics are similar (see Fig.~S2 in the Supplemental Material), indicating that 
it does not matter precisely how the stability difference between fully homochiral blocks and heterochiral ones is split into the contribution from chirality mismatched basepairs versus chirality alternation within a strand. We therefore present data only for the case $u \sim v$.

\begin{table}
\centering
\begin{tabular}{lccc}
\hline
\hline
Reaction type & Parameter & Expression & Value \rule{0pt}{3.5ex} \rule[-2.ex]{0pt}{0pt} \\
\hline 
Collision & $k_{\rm coll}$ &  & $1$ \rule{0pt}{3ex} \\
Hybridization & $k_{\rm on}$ & $\frac{1}{\theta}k_{\rm coll}$ & \\
Ligation & $k_{\rm lig}$ &  & $1.5 \times 10^{-4}$ \\
Temperature cycle & $1/\tau$ & $k_{\rm coll}\exp(\delta G_1 L_{\rm cycle})$ & $1.8 \times 10^{-3}$\\
Racemization & $k_{\rm rac}$ & & \\
Hydrolysis & $k_{\rm cut}$ & & $10^{-10}$\\
Outflow & $\phi$ & & $10^{-8}$\\
Chiral inhibition & $\sigma$ & & $0.05$ or $1$ \rule[-1.ex]{0pt}{0pt}\\
\hline
\hline
\end{tabular}
\caption{Parameters used in the simulations. Note that $\sigma$ has no units, all rates are in units of $k_{\rm coll}$ and $\tau$ is parametrized with $L_{\rm cycle}$ which is defined in equation~(\ref{eq:tau_expression}).}
\label{tab:parameter_table}
\end{table}
\subsubsection{Quantities of interest}
Without racemization of the system at the free-monomer level, the two enantiomeric populations remain constant over time. There is no interest in looking at the enantiomeric excess at the monomer level for such systems; instead we introduce the chirality parameter $\chi$, which describes the chirality of a given strand
\begin{equation}
\chi = 2 f_{\rm D} -1 \, ,
\end{equation}
where $f_{\rm D}$ is the fraction of \iupac{\D} monomers in a given strand and $\chi = 0$ describes a strand with half its monomers being \iupac{\D} chiral and the second half \iupac{\L} chiral (the enantiomeric excess of such a strand equals 0). Note that $\chi = 1$ characterizes a homochiral strand of \iupac{\D} monomers while $\chi = -1$ characterizes a homochiral strand of \iupac{\L} monomers. We also introduce a second parameter $\xi_i$, the fraction of homochiral 2-mers in the strand $i$,
\begin{equation}
\xi_i = \frac{h_i}{L_i-1} \, ,
\end{equation}
where $h_i$ is the number of homochiral \iupac{\D\D} or \iupac{\L\L} 2-mers in the strand $i$ of length $L_i$. The total fraction of homochiral 2-mers in the system $\bar{\xi}$ thus reads
\begin{equation}
\bar{\xi} = \frac{\sum_{i} \xi_i (L_i - 1)}{\sum_{i} (L_i - 1)} \, .
\end{equation}
A system composed of strands with only homochiral 2-mers will be characterized by a parameter $\bar{\xi} = 1$, and $\bar{\xi} = 0$ in the case of heterochiral 2-mers only.

\subsubsection{System without racemization}
\begin{figure*}[]
\centering
\includegraphics[scale=0.6]{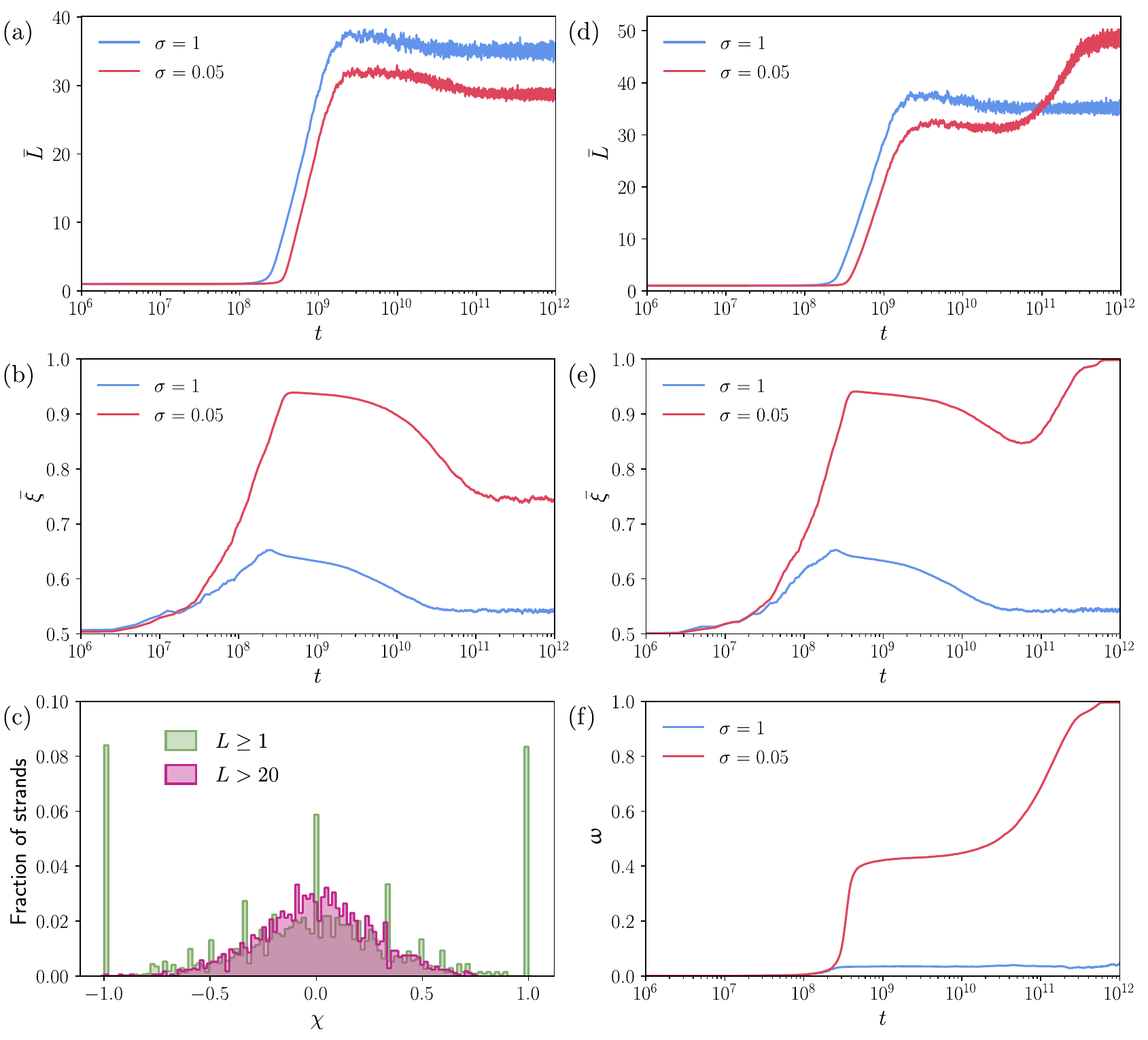}
\caption{Simulations of template-directed ligation in a closed reactor. Data are extracted from simulations performed (a)-(c) without any racemization and (d)-(f) with fast racemization. (a) Average strand length as a function of simulation time (without racemization). (b) Fraction of homochiral 2-mers $\bar{\xi}$ in the system as a function of time (without racemization). (c) Histograms of the fraction of strands in the system as a function of the chiral parameter $\chi$. These data are extracted from a single snapshot at a time step located in the steady state, without considering racemization reactions in the case $\sigma = 0.05$. The green histogram considers all strands in the system. The magenta histogram considers only strands with a length $L > 20$ monomers. (d) Average strand length as a function of simulation time in simulations, including fast racemization. (e) Fraction of homochiral 2-mers in the system as a function of time (with racemization). (f) Enantiomeric excess ($\upomega$) at the monomer level (and not only the free-monomer level) as a function of time. All data shown here are averaged over 20 independent realizations. Single trajectories are displayed in Fig.~S3 of Supplemental Material.}
\label{fig:close_reactor_graph}
\end{figure*}

Without racemization reactions, the strand average length $\bar{L}$ reaches a steady state after a transient regime of exponential growth [Fig.~\ref{fig:close_reactor_graph}(a)] which is necessarily racemic regarding its general composition. The number of monomers of both chiralities is conserved over time. However, patterns can still emerge, and we can end up with a system composed of two mirrored populations of homochiral strands. Looking at [Fig.\ref{fig:close_reactor_graph}(b)], we observe in the case with no kinetic stalling ($\sigma = 1$) that there seems to be almost no bias in the steady state of the system (the fraction of homochiral 2-mers $\bar{\xi}$ has a value close to $0.55$ which describes a system with as many heterochiral 2-mers as homochiral ones). Even though the bias emerging from the thermodynamics of RNA hybridization is mild, when we include the stalling in the simulation ($\sigma = 0.05$ here), we observe a strongly biased  steady state which contains an important fraction of homochiral 2-mers ($\bar{\xi} \sim 75\%$). This observation demonstrates that the thermodynamical bias in the energy model is unable to induce a significant bias in the final composition of the system: The presence of kinetic stalling is also required to do so (as illustrated and explained in detail in \cite{goppel_thermodynamic_2022}). As it will also be shown in the simulations with racemization, the stalling effect is a crucial component for homochirality emergence in such systems where template-directed ligation takes place.

The nonmonotonic shape of $\bar{\xi} (t)$ is explained by a multiple-step growth of strands. There is first a slight growth phase in which dimers are elongated with dimers or monomers that will consume most of the free monomers initially present in the system. Thermodynamic discrimination is strong for complexes involving short strands (such as monomers and dimers) and therefore mainly homochiral bonds will be formed during this phase. The fraction of homochiral 2-mers in the system, $\bar{\xi}$, reaches its maximal value. Once almost all monomers have been consumed in the system, ligation of longer strands will occur and the average length $\bar{L}$ will significantly grow. As strands become longer, the thermodynamic discrimination of complexes involving chiral mismatches is weakened by the long hybridized interface stabilizing the complex and replication with mismatches is eased, increasing the number of heterochiral 2-mers in the system. Eventually, the system reaches a steady state in which ligation and hydrolysis events are balanced.

Looking at the distribution of the chiral component $\chi$ in the steady state [Fig.~\ref{fig:close_reactor_graph}(c)], we observe that it is mainly centered around $\chi = 0$. Two peaks indicate a significant number of homochiral \iupac{\D} and \iupac{\L} strands, but 
only short strands contribute to these peaks. Indeed, they are only present in the green histogram that describes the whole system and disappear within the magenta histogram taking into account only strands that are $20$ monomers long and more. Therefore, the natural thermodynamic bias plus the inherent stalling of template-directed ligation of RNA is not sufficient to produce a pool of homochiral polymers only, even though a significant portion of the strands will present a chiral bias with an $\upomega \geq 50\%$.

\subsubsection{Simulations with racemization}
We now describe simulations with racemization. In this section, we assume  racemization to be infinitely fast. In practice, this means that we equilibrate  the numbers of \iupac{\D} and \iupac{\L} monomers after each reaction which affects these numbers. In Sec.~\ref{section:effects_of_racemization}, we reconsider this assumption of infinitely fast racemization, and instead treat racemization as a standard chemical reaction with a characteristic time that will be varied.

The important new feature introduced by these simulations with racemization is the existence of  a second growth phase for the average length $\bar{L}$ when kinetic stalling is present ($\sigma = 0.05$) during ligation events [Fig.~\ref{fig:close_reactor_graph}(d)]. The mean length reaches first a transient plateau (corresponding to the steady-state of the experiment without racemization) and then undergoes a second growth phase towards a final stationary state. This second growth phase is explained by the increase of one enantiomer population at the expense of the other one. Indeed, $\upomega$ [Fig.~\ref{fig:close_reactor_graph}(f)] increases significantly in the racemized experiment and eventually reaches $\upomega\sim~ 100\%$, which corresponds to a completely homochiral system. While the system is converging towards a single chiral state, ligation events are less and less stalled since chiral mismatches at ligation sites are less likely to occur as the chiral homogeneity of the system grows. We believe that this second growth could be particularly relevant for the emergence of life, as discussed in Sec.~\ref{sec:discussion}. Comparing simulations with and without racemization, we observe that the case without chiral stalling ($\sigma = 1$) gives similar results. With racemization but no stalling, the system reaches an $\upomega \sim 4-5\%$ due only to the thermodynamic discrimination of heterochiral phosphodiester bonds or base pairs during strand hybridization. In the case of chiral stalling (observed experimentally by Joyce {\it et al.} \cite{joyce_chiral_1984} and Bolli {\it et al.} \cite{bolli_pyranosyl-rna_1997}) adding racemization induces a convergence towards a total homochiral system.
In this case, the average fraction of homochiral 2-mers converges to $\bar{\xi} = 1$ [Fig.~\ref{fig:close_reactor_graph}(e)].

\subsubsection{Effects of racemization reaction speed}
\label{section:effects_of_racemization}
To ensure that the observed results are not specific to the particular case of instantaneous racemization for non polymerized monomers, we investigate the effects of the racemization reaction speed on the enantiomeric excess. We model this reaction assuming the mass-action law, with $k_{\rm rac}$ the rate constant for the reversible conversion reaction $\iupac{\D} \rightleftharpoons \iupac{\L}$ at the free-monomer level.

\begin{figure}[h]
\centering
\includegraphics[scale=0.6]{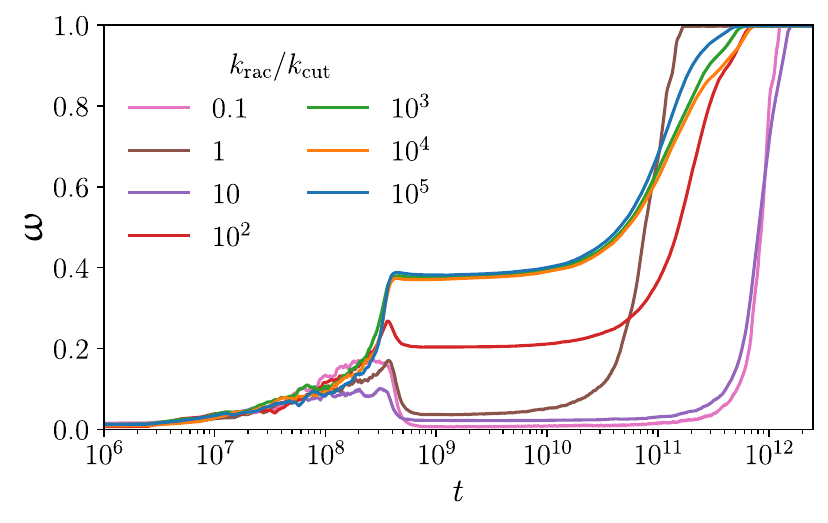}
\caption{Enantiomeric excess as a function of time in a template-directed ligating closed system in the presence of chiral stalling $\sigma = 0.05$. The enantiomeric excess is computed over all polymerized monomers. Simulations were carried out with $k_{\rm cut} = 10^{-10}$, however, for computational time purposes, brown and pink curves were simulated with $k_{\rm cut} = 10^{-9}$ (which explains the difference in timescales for the transition to occur). Data shown here are averaged over 20 independent realizations starting from an initial racemic pool of 5000 nucleotides.}
\label{fig:close_reactor_diff_racemization}
\end{figure}
Figure~\ref{fig:close_reactor_diff_racemization} shows first that the enantiomeric excess saturates at a limit value when the racemization rate constant exceeds $k_{\rm rac}~=~10^{-7}$ (in units of $k_{\rm coll}$) and also that the system always converges towards a homochiral state, even when $k_{\rm rac} < k_{\rm cut}$, which means that the racemization reaction has the slowest rate constant of the system. However, we observe that the timescale for the system to converge increases as the racemization slows down, and it also affects the height of the transient plateau in the simulation. This makes sense as for rapid racemization, a slight imbalance of polymerized nucleotides appears stochastically and gets amplified, while the system is racemized at the monomer level. When racemization slows down, non-polymerized monomers are not racemized, and strands are formed out of the chirality that dominates free monomers. Looking at the enantiomeric excess for non polymerized monomers, this effect tends to decrease the chiral asymmetry before a transient plateau is reached. The influence of temperature cycles and of the chiral stalling amplitude is also studied in Figs.~\ref{fig:close_reactor_diff_temperature} and \ref{fig:close_reactor_diff_stalling} of the Appendix.

\subsection{Chiral symmetry breaking in an open reactor}
\label{subsec:open_reactor}
\subsubsection{Injection of a racemic mixture}
By open reactor we mean a reactor, where matter can flow in and out of the system. A well-known example of an open reactor is a continuously stirred tank reactor, in which species are supplied from the environment at a certain rate and then species in excess flow out of the system at the same rate after reaction. This rate is denoted by $\phi$ below and its inverse represents the residence time of species in the system. Furthermore, hydrolysis is neglected in this section. 

Here we assume specifically that the injected mixture is composed of only monomers and dimers and that this injected mixture is racemic. For this reason, the symmetry is not explicitly broken by the injection of matter from the environment, similarly to Ref.~\cite{laurent_emergence_2021}, and we are interested in a spontaneous chiral symmetry breaking. We also assume that monomers and dimers reach a stationary state in the open reactor more quickly than other species, so their concentrations can be assumed to be fixed while the concentrations of other species are allowed to vary in time. 

In practice, the system is inoculated with a total of 2800 nucleotides distributed in 2000 monomers (1000 \iupac{\D} and 1000 \iupac{\L} monomers) and 400 dimers (100 \iupac{\D\D}, 100 \iupac{\L\L}, 100 \iupac{\D\L} and 100 \iupac{\L\D} dimers) in a racemic fashion. In this experiment, we investigate whether we can reach a homochiral state without the racemization of monomers.

\begin{figure*}[]
\centering
\includegraphics[scale=0.6]{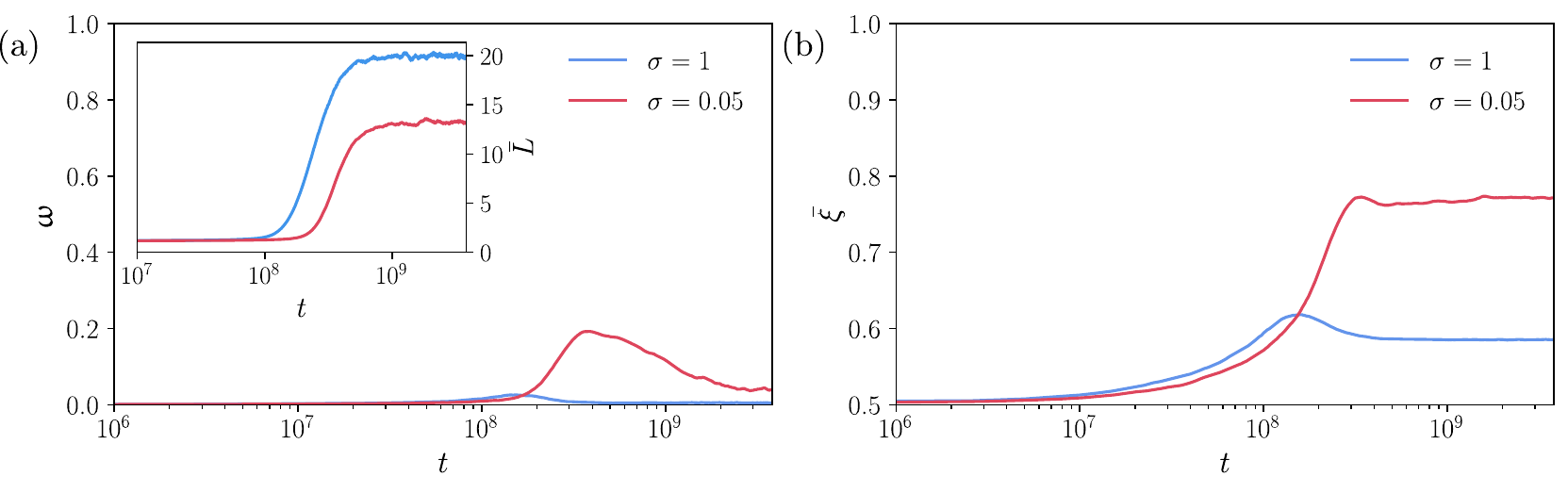}
\caption{Template-directed ligation in an open reactor in the absence of racemization reaction. (a) Enantiomeric excess at the monomer level as a function of time. The inset shows the average length in the system over time. (b) Fraction of homochiral 2-mers $\bar{\xi}$ in the system as a function of time. All data shown here are averaged over 100 independent realizations. Single trajectories are displayed in Fig.~S4 of Supplemental Material.}
\label{fig:open_reactor_graph}
\end{figure*}

It appears first that the steady state reached by the system is not fully homochiral [Fig.\ref{fig:open_reactor_graph}(a)]. In the case with chiral stalling, the final $\upomega$ is low (approximately $4-5\%$). The fraction of homochiral 2-mers reaches a relatively high stationary value of $\bar{\xi} \sim 75\%$, which however does not imply homochirality at the strand level. Indeed, a strand composed $50\%$ of \iupac{\D} monomers and $50\%$ \iupac{\L} monomers could still have only one heterochiral 2-mer while being perfectly heterochiral ($\chi = 0$). Initially, due to statistical fluctuations, an excess of short strands using preferentially one chirality will be produced. They will start to produce their kind autocatalytically, and thus one of the two populations will contain longer and longer strands. This explains the increase in $\upomega$ observed in Fig.\ref{fig:open_reactor_graph}(a). Unlike with the closed reactor case, here there is no conservation law for monomers that would slow the growth of strands of one chirality. Strands mainly composed of one chirality quickly reach a steady state for their average length. Since no racemization is present in the system, there is no real negative interaction between monomers of different chirality (except during hybridization and ligation) and the two populations are essentially independent. Therefore, strands composed of the  non dominant chirality will start to grow, and their concentration will exhibit a similar but retarded growth phase as the dominant chirality. Eventually, the initially smaller population will catch up with the dominant one, causing the vanishing of the enantiomeric excess. In any case, the fraction of homochiral 2-mers in the system remains high [Fig.\ref{fig:open_reactor_graph}(b)] in the steady state. 

\subsubsection{Injection of a biased mixture}
So far, we have assumed the injected mixture to be racemic, which means that the chiral symmetry of the system is not broken explicitly. Now, we will consider instead a scenario with explicit symmetry breaking. This could happen if the solution which is injected in the system is not itself racemic because it involved, for instance, the synthesis of riboses which was not symmetric. It has been shown, for instance, that the synthesis of pentoses catalyzed by homochiral \iupac{\L\L-dipeptides} in prebiotic conditions is not symmetric and produces in particular a greater amount of \iupac{\D-riboses} than its enantiomer \cite{pizzarello_stereoselective_2010}. In a closed reactor, as demonstrated in Sec.~\ref{sec:results_closed_reactor}, an initial racemic system converges towards a homochiral state.  In the case of the open reactor without racemization reactions, we saw that even though a transient state with a significant $\upomega$ is achieved, the steady state lies around the racemic state. Therefore, we simulated initially biased open reactors to study how the bias is amplified by the two different types of polymerization. 
\begin{figure}
\centering
\includegraphics[scale=0.85]{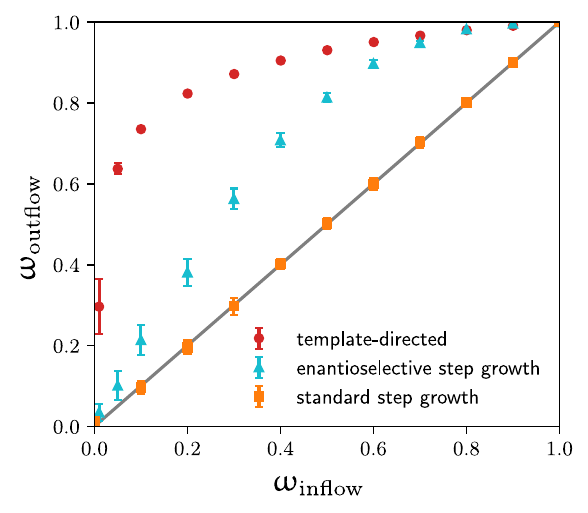}
\caption{Final enantiomeric excess as a function of the enantiomeric excess of the injected mixture in the open reactor. In the simulation, the injected mixture contains 2800 nucleotides except for the standard step-growth simulations where it contains  only 700 nucleotides. Chiral stalling is set to $\sigma = 0.05$ in the template-directed experiment. The enantioselective experiment has a bias $\alpha = 10^{-10}$ and the standard one has no bias. The outflow rate is $\phi = 10^{-8}$ and the ligation rate for the step-growth polymerization is $k_{\rm lig} = 5.27\times 10^{-9}$. Examples of single trajectories are displayed in Fig.~S4 of Supplemental Material.}
\label{fig:biased_amplification}
\end{figure}
It turns out, as one could expect, that unbiased step-growth polymerization conserves the enantiomeric excess of the injected mixture (results are gathered in Fig.~\ref{fig:biased_amplification}). The bias is amplified in the case of a perfectly chiroselective step-growth process. Still, this mechanism's lack of autocatalytic feedback means that the amplification  is low compared to the case of templated ligation. Indeed, template-directed ligation in the presence of chiral stalling (here $\sigma = 0.05$) amplifies best the bias present in the injected mixture, especially when it has a low value: for instance a bias of $\upomega=10\%$ can be boosted to values as high as $\upomega=75\%$ without the need of any racemization reactions. These results change depending on the concentration of the injected mixture or ligation rates and should be considered indicative only. Still, it should be noted that the selective step-growth polymerization results do not vary much when changing the concentration of the chemostat or the ligation rate.

\section{Discussion}
\label{sec:discussion}
The main result of the closed reactor simulations is that RNA template-directed ligation with racemization reactions at the monomer level can converge toward a homochiral system of RNA polymers, in contrast to step-growth polymerization which can only lead to a rather small degree of homochirality.
The kinetic chiral selection during template-directed ligation observed by Joyce {\it et al.} \cite{joyce_chiral_1984} is a crucial component of the mechanism because without it the system remains close to the racemic state (Fig.~\ref{fig:close_reactor_graph}). Effectively, this kinetic selection implements a chiral cross-inhibition process. A conceptually similar process occurs experimentally in a chiroselective peptide replicator \cite{saghatelian_chiroselective_2001}, in which autocatalytic amplification of homochiral compounds is reported, in contrast to the non autocatalytic production of heterochiral compounds. The former occurs by ligation on a template and is fast, whereas the latter is much slower and does not require a template.

In our model, one of the two chiralities is removed from the system, but which chirality disappears is random: A minute imbalance due to the stochasticity of chemical reactions can be amplified, such that one chirality dominates over the other one.
In the final homochiral steady state, heterochiral strands and homochiral strands of the chirality that has been erased are no longer formed, which implies a reduction of the chemical space available to molecules in this system. This reduction of the chemical space is remarkable because it is unfavorable from a thermodynamic point of view. Indeed, homochiral strands have a lower entropy than heterochiral strands and should be favored thermodynamically. Here the production of lower-entropy species is only possible because the system is far from equilibrium, so this loss of entropy can be compensated for by the spending of sufficient free energy. 

It is important to emphasize that the path to homochirality of RNA described here differs fundamentally from a scenario where homochirality would be established already at the level of monomer production, e.g., with a mechanism akin to the Frank model. Instead, in our scenario, the formation of RNA strands in a non equilibrium interplay between hybridization, ligation, and cleavage effectively acts as a chirality filter, which purifies the overall chirality of the RNA strands. Clearly, the action of this filter would also increase the chances for the emergence of ribozymes that might reinforce homochirality, at the level of oligomers, monomers, or both.  

Another aspect of our scenario here is that together with the disappearance of one chirality, the average length of strands in the homochiral steady state increases [Fig.~\ref{fig:close_reactor_graph}(d) and \ref{fig:close_reactor_graph}(e)]. This occurs naturally as more nucleotides from the dominant chirality become available to join in template-directed ligation without chiral stalling. This effect has significant consequences: The reduction of the chemical space benefits the length of the strands in the system, which is important for the emergence of life, where the formation of long RNA is a bottleneck \cite{mast_escalation_2013}.

We also explore the role of racemization reactions at the monomer level. In particular, we find that these reactions do not need to be very fast. Regarding the open reactor simulation with chiral stalling and no racemization, even though no significant $\upomega$ is reached during the steady state, we observe an interesting transient behavior when $\upomega$ reaches a maximum value. At this point, one chirality is much more present in the system, and a large part of the system is composed of monomers of this dominant chirality. In addition, a large proportion of those strands have an $\upomega > 50 \%$. From this transient state a first ribozyme composed of the dominant chirality could emerge, accelerating the replication of its kind \cite{horning_amplification_2016} and thus amplifying definitely the enantiomeric excess and eventually driving the system towards homochirality. In addition, template-directed ligation appears to be a good amplification mechanism of preexisting chiral biases in open systems, especially when these biases are low. 

\section{Conclusion}
\label{sec:conclusion}
In this article, we have shown that a closed system initially inoculated with a racemic mixture of RNA monomers and dimers, performing template-directed ligation converges towards a homochiral state in the presence of racemization reactions and chiral stalling. Another important ingredient in our model is that single-stranded polymers can be broken into smaller oligomers or monomers thanks to hydrolysis in contrast to duplexes which are protected against hydrolysis. As a result, monomers and oligomers are recycled and due to racemization reactions, no chiral bias can build up at the monomer level, which allows us to reach the maximal level of homochirality in the polymers. 

This mechanism is robust because there is no need to fine tune kinetic parameters such as parameters associated with the level of chiral selectivity, the specificity or the kinetics of the racemization reactions.  
Overall, the route we propose is satisfying as it is compatible with the RNA world scenario \cite{higgs_rna_2015} and with studies on the amplification of homochirality in peptides  \cite{saghatelian_chiroselective_2001}.

Our detailed model also reinforces recent studies \cite{chen_origin_2020, tupper_role_2017} that homochirality could have emerged alongside as opposed to before or after the first functional polymers due to template-directed polymerization. In addition, template-directed ligation of RNA has recently been studied as an efficient mechanism to produce polymeric homochirality at the duplex level \cite{ross_duplex_2022}, starting from a racemic pool of nucleotides. Joyce {\it et al.} \cite{joyce_chiral_1984}, puzzled by their discovery, stated about the chiral stalling that ``inhibition raises an important problem for many theories of the origin of life." Almost 40 years later, it appears that chiral stalling is instead the phenomenon that makes RNA template-directed ligation an interesting candidate for the emergence of homochirality in a prebiotic RNA world.

\section{Acknowledgments}

We acknowledge stimulating discussions with J. Ribo, O. Trapp, J. Rosenberger and B. Altaner. This work was supported by the German Research Foundation (DFG) through U.G., via the TRR 235 Emergence of Life (Project No. 364653263) and the excellence cluster ORIGINS. D.L. received support through Grants No. ANR-11-LABX-0038 and No. ANR-10-IDEX-0001-02.

\appendix
\section{Robustness of the transition towards homochirality}
\label{robustess_of_the_transition}
\subsection{Effects of temperature cycles}
\label{sec:temperature_effects}
The influence of the period of temperature cycles on the dynamics of the systems has also been studied. Temperature cycles are characterized by the time period $\tau$ parametrized by the length $L_{\rm cycle}$ such that 
\be
\tau =\frac{1}{k_{\rm coll}}\exp (-\delta G_1 L_{\rm cycle}) \, .
\label{eq:tau_expression}
\ee
This choice of parametrization means that on average, duplexes formed by strands of length $ l < L_{\rm cycle}$ will have time to dehybridize between two temperature peaks while duplexes formed by strands of length $l > L_{\rm cycle}$  will not and are for the most part, melted during the temperature elevation, i.e. the higher the $L_{\rm cycle}$, the longer the time for duplexes to be dehybridized.
\begin{figure}[h]
	\centering
	\includegraphics[scale=0.57]{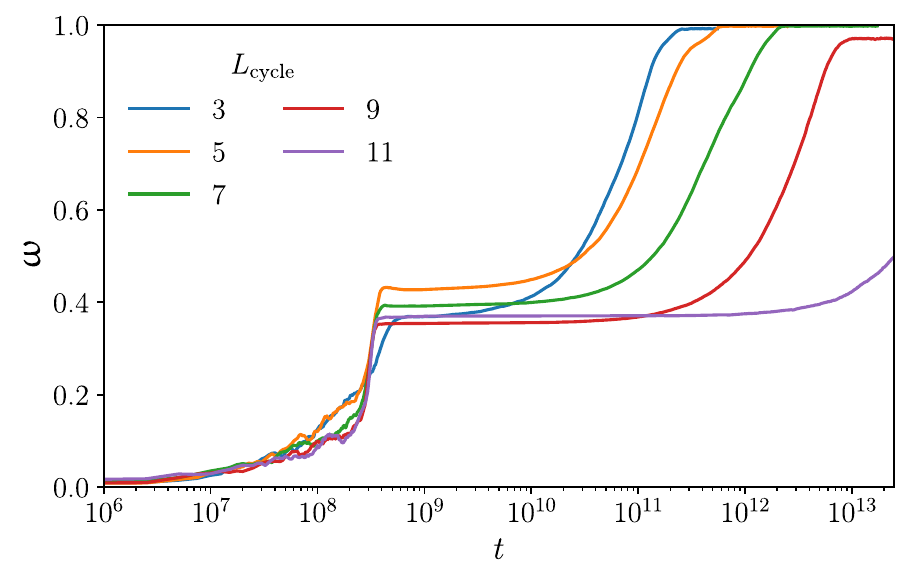}
	\caption[Enantiomeric excess as a function of time in a template-directed ligating closed system for different temperature cycles]{Enantiomeric excess as a function of time in a template-directed ligating closed system in the presence of chiral stalling $\sigma = 0.05$ for different temperature cycles. The enantiomeric excess is computed over polymerized monomers only. Data shown here are averaged over 20 independent realizations starting from an initial racemic pool of 5,000 nucleotides with instantaneous racemization for non polymerized monomers.}
	\label{fig:close_reactor_diff_temperature}
\end{figure}
This phenomenon can be observed in Fig.~\ref{fig:close_reactor_diff_temperature}, where the dynamics spans over more orders of magnitude of time as the length $L_{\rm cycle}$ increases and thus the time $\tau$ between two temperature peaks increases (note that $\delta G_1 < 0$). With the increase of $\tau$, the dynamics slows down but the system still end up in a homochiral state in the end. However, we notice that for typical lengths $L_{\rm cycle} > 7$, the degree of homochirality of the steady state starts to slightly decrease. This could be due to the fact that hybridized strands are protected against cleavage as hydrolysis is only possible for single-stranded RNA. This induces the protection of the remaining monomers of the disappearing chirality against the conversion mechanism that would require first the hydrolysis of a strand resulting in the production of a free monomer that ends up converted to the other chirality, because strands remain hybridized for longer times as $L_{\rm cycles}$ increases. Eventually, in the limit $L_{\rm cycle} \to \infty$ and thus $\tau \to \infty$, the dynamics would end up frozen before any state with a significant $\upomega$ would appear, as duplexes formed by long strands would be highly stable [see Eq.~(\ref{thermo_consist})] and the typical time to separate the duplex would be too large to see the system evolve on a reasonable timescale.

\subsection{Effects of the chiral stalling amplitude}
In this section, we generally used the reference value $\sigma = 0.05$ for the chiral stalling as it corresponds to a reduction between one and two order of magnitude of the ligation speed when chiral mismatches occur around ligation sites and is expected from experiments. We investigated the effect of changes in the chiral stalling amplitude and found that it affects both the timescales of the dynamics and the final stationary state.
\begin{figure}[h]
	\centering
	\includegraphics[scale=0.57]{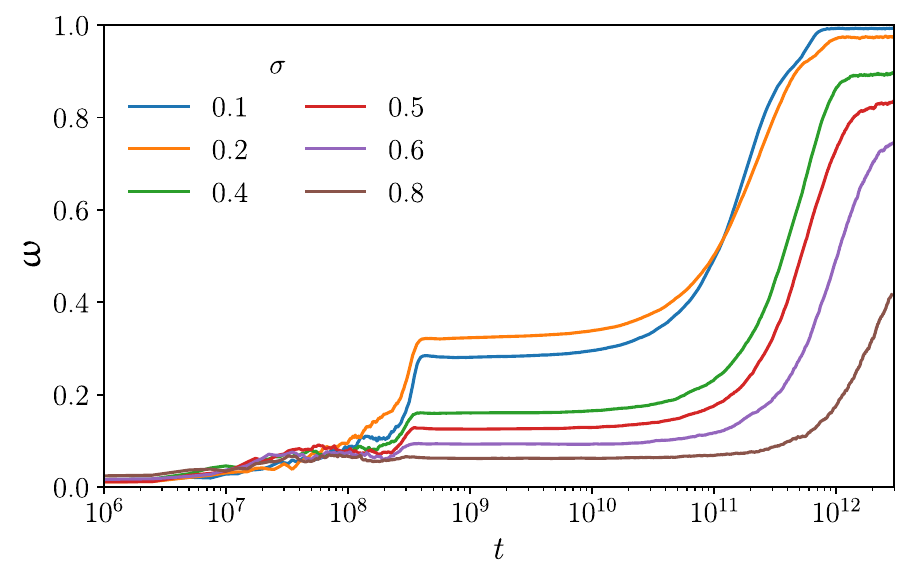}
	\caption[Enantiomeric excess as a function time in a template-directed ligating closed system for different kinetic stalling]{Enantiomeric excess as a function of time in a template-directed ligating closed system for different kinetic stallings. The enantiomeric excess is computed over polymerized monomers only. Data shown here are averaged over 20 independent realizations starting from an initial racemic pool of 5,000 nucleotides with instantaneous racemization for non-polymerized monomers.}
	\label{fig:close_reactor_diff_stalling}
\end{figure}
Indeed, Fig.~\ref{fig:close_reactor_diff_stalling} shows that for chiral stalling with values above 0.2, the system does not end up in a full homochiral state but in a partial one, with enantiomeric excesses that decrease as $\sigma$ gets closer to 1. Eventually, for $\sigma = 1$, the system reaches an almost racemic steady state [see Fig.~\ref{fig:close_reactor_graph}(f)]. Looking at the graph, it is unclear if these partially homochiral stationary states are really steady states 
or if it is just that the time required for the changes in $\upomega$ gets longer for high values of $\sigma$. However, it is important to note that the logarithmic scale compresses the right part of the plot associated with long times $t$. With a closer look, we observe that the simulations indeed reach steady states (except for $\sigma = 0.8$, which did not have the time to converge). In any case, we observe that there seems to be a threshold above which no significant enantiomeric excess can appear in a reasonable time. We also can be confident with the value $\sigma = 0.05$ used in our study as the stalling observed by Joyce {\it et al.} \cite{joyce_chiral_1984} and Bolli {\it et al.} \cite{bolli_pyranosyl-rna_1997} was strong in both cases. Also the case $\sigma = 0.2$ gives results similar to the ones found earlier and it corresponds to an $80\%$ reduction of the ligation speed.

\end{document}